\documentclass[usenatbib]{mnras} 
\usepackage{graphicx, amssymb, natbib, bm}

\usepackage{mathptmx}
\usepackage{epsfig}
\usepackage{wrapfig}
\usepackage{amsmath}
\usepackage{natbib}
\usepackage{graphicx}
\usepackage{subfigure}
\usepackage{enumitem}
\usepackage{breqn}

\newcommand\msun{\hbox{\ensuremath{{\rm M}_{\odot}}}}

\newcommand{\scinot}[1]{\ensuremath{\times 10^{#1}}}

\newcommand{\cse}{Computing in Science \& Engineering}

\title[GCs in UDGs]{An Excess of Globular Clusters in UDGs Formed Through Tidal Heating}
\author[Carleton et al.]
{Timothy Carleton,$^{1}$\thanks{$\!\!$e-mail: carletont@missouri.edu}
	Yicheng Guo$^1$, Ferah Munshi$^2$, Michael Tremmel$^3$, Anna Wright$^4$\\
	$\!\!^{1}$Department of Physics and Astronomy, 223 Physics Building, University of Missouri, Columbia, MO 65211, USA \\	
	$^2$Department of Physics \& Astronomy, University of Oklahoma, 440 W. Brooks St., Norman, OK 73019, USA\\
	$^3$Yale Center for Astronomy \& Astrophysics, Physics Department, P.O. Box 208120, New Haven, CT 06520, USA\\
	$^4$Department of Physics \& Astronomy, Rutgers, The State University of New Jersey, 136 Frelinghuysen Road, Piscataway, NJ 08854, USA\\
}

\begin{document}
	\pagerange{\pageref{firstpage}--\pageref{lastpage}} 
	\pubyear{2020}
	\maketitle
	\begin{abstract}
To investigate the origin of elevated globular cluster abundances observed around Ultra-Diffuse Galaxies (UDGs), we simulate globular cluster populations hosted by UDGs formed through tidal heating. Specifically, globular cluster (GC) formation is modeled as occurring in regions of dense star formation. Because star-formation-rate-densities are higher at high redshift, dwarf galaxies in massive galaxy clusters, which formed most of their stars at high redshift, form a large fraction of their stars in globular clusters. Given that UDGs formed through environmental processes are more likely to be accreted at high redshift, these systems have more GCs than non-UDGs. In particular, our model predicts that massive UDGs have twice the GC mass of non-UDGs of similar stellar mass, in rough agreement with observations. Although this effect is somewhat diminished by GC disruption, we find that the relationship between GC mass fraction and cluster-centric distance, and the relationship between GC mass fraction and galaxy half-light radius are remarkably similar to observations. Among our model objects, both UDGs and non-UDGs present a correlation between halo mass and GC mass, although UDGs have lower dynamical masses at a given GC mass. Furthermore, because of the effectiveness of GC disruption, we predict that GCs around UDGs should have a more top heavy mass function than GCs around non-UDGs. This analysis suggests that dwarfs with older stellar populations, such as UDGs, should have higher globular cluster mass fractions than objects with young stellar populations, such as isolated dwarfs.
	\end{abstract}

\begin{keywords}
	galaxies: formation, evolution, dwarf, haloes, clusters, star clusters
\end{keywords}

\section{Introduction}
The recent identification of a large population of Ultra-Diffuse Galaxies (UDGs) in clusters \citep{vandokkum2015} has generated significant interest in Low-Surface-Brightness Galaxies. These dwarf galaxies are characterized by stellar masses ranging from $10^{7}-10^9$~\msun{} and half-light radii extending from $1.5$ to $7$~kpc \citep{koda2015,yagi2016}. Observations indicate that, {although a population of UDGs is present in low-mass groups and the field \citep{roman2017,leisman2017}}, most UDGs are found in cluster environments \citep{vanderburg2017}, and, like other dwarfs in clusters, are characterized by old stellar populations and low metallicities \citep{ferre-mateu2018}.

{Generally, theories for UDG formation describe them as dwarf galaxies that have been enlarged due to internal or external processes. For example, \cite{amorisco2016} suggested that UDGs represent galaxies living in halos in the high-spin tail of the spin distribution, with their large sizes the result of high angular-momentum gas. Alternatively, strong feedback from supernovae and stellar winds has been shown to increase the sizes of dwarf galaxies in simulations to resemble UDGs \citep{dicintio2017,chan2018}. In a similar vein, it has been suggested that the period of globular cluster formation early in the Universe was violent enough in some systems to completely eject their gas, leaving a galaxy with a low stellar mass and large size, but a large halo mass and GC population \citep{agertz2016,vandokkum2016}. While these models presume that UDGs in clusters have early infall times to prevent re-accretion of gas, other models suggest a more explicit environmental mechanism. For example, ram-pressure stripping or tidal heating may puff up galaxies \citep{yozin2015,safarzadeh2017,ogiya2018}. In particular, \cite{carleton2019} modeled UDGs as tidally-heated dwarfs in clusters, and reproduced many observed UDG properties, such as their size distribution and old stellar populations.
Lastly, some comprehensive simulations suggest that a combination of these effects may be at play \citep{jiang2019, liao2019,tremmel2019,martin2019,sales2020,wright2020}.}

Observations indicate that the environments in which UDGs are found have a substantial impact on their formation and evolution. The relative abundance of UDGs in a cluster is dependent on the halo mass of the cluster \citep{vanderburg2017}, and the morphologies of UDGs appear to evolve from disks in the field to elongated spheroids in clusters \citep{burkert2017,rong2019}. Furthermore, although a population of UDGs is observed in the field \citep{williams2016,leisman2017,roman2017iso}, they have different properties than UDGs observed in clusters \citep{prole2019b}, so they may have a distinct formation process than cluster UDGs.
However, environmental processes have generally been unable to explain one of the most intriguing aspects of UDGs: their unusual globular cluster (GC) populations.
Multiple studies \citep{vandokkum2017,lim2018,amorisco2018,prole2019} have confirmed that a substantial number of UDGs {in clusters} host exceptionally large GC populations, with $>2$ times higher GC abundances than non-UDG dwarfs at a similar stellar mass. 
Further evidence of the unusual GC populations hosted by UDGs comes from \cite{vandokkum2018}, which found that GCs around the UDG DF2 have an unusually top-heavy luminosity function. {However, some observations indicate that UDGs in less dense environments don't show the same elevated GC abundances as UDGs in clusters \citep{somalwar2020}, suggesting that the cluster environment plays a role in evolution of their GC populations.} All of this evidence, along with the unusually high velocity dispersions observed in some UDGs \citep{vandokkum2016}, has led some to speculate that UDGs in clusters live in over-massive dark-matter halos.

Despite their use as probes of dark-matter halos, there remains significant uncertainty regarding how globular clusters are formed. Their old ages and low metallicities suggest that they primarily form at very high redshift ($z>5$; \citealt{forbes2010,boylan-kolchin2017}). On the other hand, the mixed stellar populations of some globular clusters \citep{gratton2012}, as well as the continued formation of star clusters suggests that globular clusters can be formed and destroyed throughout cosmic time. In particular, the regions of dense star formation at $z=1-2$ \citep{guo2015,elmegreen2018} may be the progenitors of some of today's globular clusters \citep{kruijssen2015}.

Notably, observations suggest that the fraction of stars within a gas cloud that form bound clusters (otherwise known as the cluster formation efficiency, $\Gamma$) is proportional to the star-formation rate surface density of the gas \citep{goddard2010,li2018}. {Previous work has recognized that a consequence of this correlation is that dwarf galaxies that fall into a cluster early (and form most of their stars in high star-formation rate density regions at high $z$) may be expected to have larger globular cluster populations \citep{mistani2016,pfeffer2018}.}
If UDGs were formed at earlier times than typical dwarf galaxies (as expected by environmental formation models like \citealt{carleton2019}, \citealt{yozin2015}, or \citealt{tremmel2019} and confirmed by observations from \citealt{ferre-mateu2018} and \citealt{ruiz-lara2018}), it is very likely they formed a higher fraction of their stellar mass in clusters. This offers the possibility that the large globular cluster populations observed in UDGs are a consequence of their early formation. Importantly, this scenario is possible for any model in which UDGs are formed at earlier times than non-UDGs and globular clusters are formed more efficiently at high $z$.

In this paper, we elaborate on the model of \cite{carleton2019} to explore the formation of globular clusters in UDGs formed through tidal heating. In particular, we explore the possibility that UDGs have large GC populations because they formed their stellar mass earlier than non-UDGs. In Section~\ref{sec:model} we describe our model of UDG formation and their associated GCs. In Section~\ref{sec:results} we describe how the GC populations of UDGs and non-UDGs compare, and how they compare with observations. In Section~\ref{sec:modellimitations} we discuss the limitations of our model when considering low-mass galaxies, and in Section~\ref{sec:conclusions} we summarize our conclusions. Throughout, we assume a $\Lambda$CDM cosmology based on the \cite{planck2016} cosmological parameters: $H_0=67.74$~km~s$^{-1}$~Mpc$^{-1}$, ${\rm \Omega_m}=0.3089$, and ${\rm \Omega_\Lambda}=0.6911$.

\begin{figure}
	\centering
	\includegraphics[width=1\linewidth]{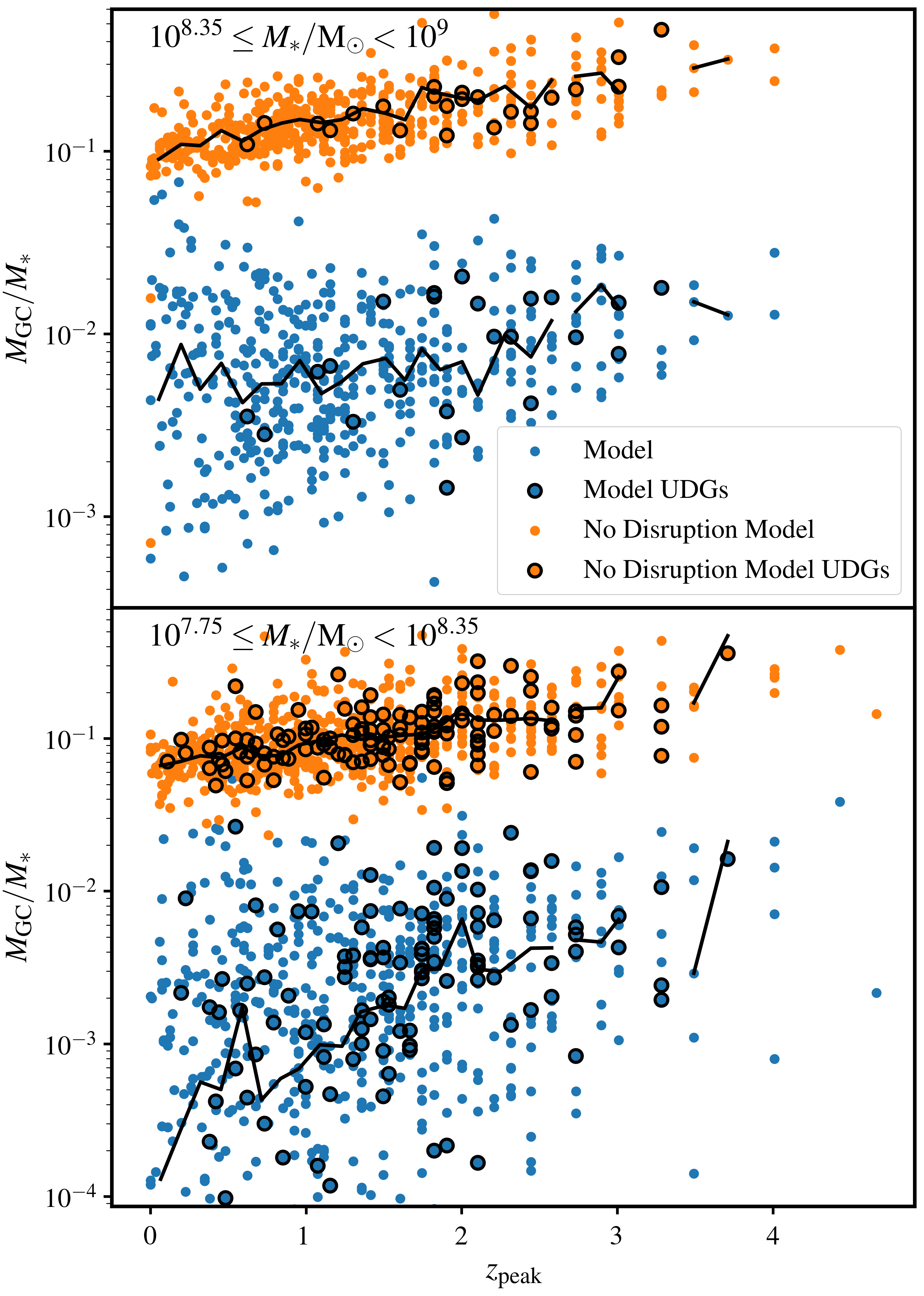}
	\caption{The relationship between GC mass fraction and infall redshift for objects in our model, split into two bins of stellar mass. Our fiducial model is shown as blue points, whereas a model with no GC disruption is shown with orange points. Points outlined in black highlight objects identified as UDGs. Because star formation is more dense at higher redshift, there is a positive correlation between GC fraction and infall redshift in both models: the black lines show the median-binned trend for both samples. Adding disruption substantially increases the scatter in this relationship, but results in more realistic GC fractions.}
	\label{fig:fracgcvszpeakbymass812m1}
\end{figure}

\section{Model}
\label{sec:model}
{To investigate the formation of GCs around {cluster} UDGs produced through tidal heating, we apply the model for tidal stripping described in \cite{carleton2019} to the Illustris TNG-100 simulation \citep{springel2018,naiman2018,marinacci2018,pillepich2018,nelson2018}.} From the Illustris-TNG Simulation, galaxies are selected from within $R_{200}$ of a massive cluster with $M_{200}>2\times10^{14}$~\msun{} at $z=0$, where $M_{200}=200\rho_{\rm crit}\frac{4\pi}{3}R_{200}^3$ and $\rho_{\rm crit}$ is the critical density of the Universe. The orbits of satellites are tracked throughout their time in the cluster. For each satellite in TNG, the infall stellar mass is taken as the stellar mass of the simulated galaxy at the time of the peak halo mass. Each galaxy is assigned a stellar half-light radius ($r_e$) based on the size-mass relation for red galaxies from \cite{lange2015}, and its halo is modeled with a cored profile, with a concentration assigned following the mass-concentration relation from \cite{prada2012}.\footnote{Although stellar sizes and halo concentrations are available for individual objects in the simulation, we use model parameters in order to focus specifically on UDGs produced through tidal heating as in the \cite{carleton2019} model. The comparatively large number of galaxies in the simulation would result in a large number of UDGs produced through internal processes, which are not the focus of this work.} At each pericentric passage, the mass within the tidal radius of the subhalo is used as input to tidal tracks of \cite{penarrubia2010} to evolve the $V_{\rm max}$ and $r_{\rm max}$ of the subhalo. The mass within $r_{\rm max}$ of the subhalo at $z=0$ compared with the mass within $r_{\rm max}$ at infall is used as input for the tracks of \cite{errani2018} to determine the amount of stellar mass loss and the change in $r_{e}$. In this work, we assign all galaxies cored dark-matter halos, which are expected to be the hosts of UDGs produced through tidal heating. As in \cite{carleton2019}, we alter the central slope of the dark-matter halo in baryon-dominated galaxies where the stellar mass within the half-light radius is higher than the dark-matter mass within the half-light radius. This procedure has been able to reproduce many aspects of the UDG population, including the size distribution and the old ages of UDGs.

By analyzing the Illustris-TNG simulation with this procedure, we have access to the cold gas and star-formation-rate properties of dwarf galaxies before and during infall, which can be used to model their globular cluster populations. {To generate model globular-cluster populations among our model dwarf galaxies, we reference the procedure of \cite{mistani2016}.} This model is based on the observation that the fraction of stars formed in clusters is proportional to the surface density of star formation. From each snapshot of each dwarf galaxy considered, we identify the average surface density of star formation of each gas particle in the dwarfs in our sample as $\Sigma_{\rm SFR}=\frac{\rm SFR_3}{\pi r_{3}^2}$, where ${\rm SFR_3}$ is the sum of the star formation rates of the $3$ nearest star forming gas particles, and $r_3$ is the distance to the third nearest star-forming gas particle. Following \cite{goddard2010}, we take the fraction of stars formed in clusters ($\Gamma$) to be 
\begin{equation}
\Gamma=0.29 \left(\frac{\Sigma_{\rm SFR}}{\msun{}~{\rm yr^{-1}~kpc^{-2}}}\right)^{0.24}.
\end{equation}
The total globular cluster mass is correspondingly increased by $\Gamma{\rm SFR}\Delta t$, where $\Delta t$ is the time between timesteps in the simulation and SFR is the total star formation rate of the galaxy. Given the mass formed in globular clusters, we populate the globular cluster mass function of \cite{jordan2007}. Each globular cluster is also assigned a position within its host galaxy following a \cite{plummer1911} distribution with a scale radius $1.5$ times the half-light radius of the galaxy. This assumed distribution is motivated by observations indicating that the extent of the GC population around UDGs is similar to the stellar extent \citep{amorisco2018}. {While the GC population is much more extended around massive galaxies like the Milky Way, their GCs are primarily built up from accretion of GCs from dwarf galaxies \citep{mackey2004,beasley2018}, so it is not expected that UDGs would have substantially extended GC populations.}

\begin{figure}
	\centering
	\includegraphics[width=1\linewidth]{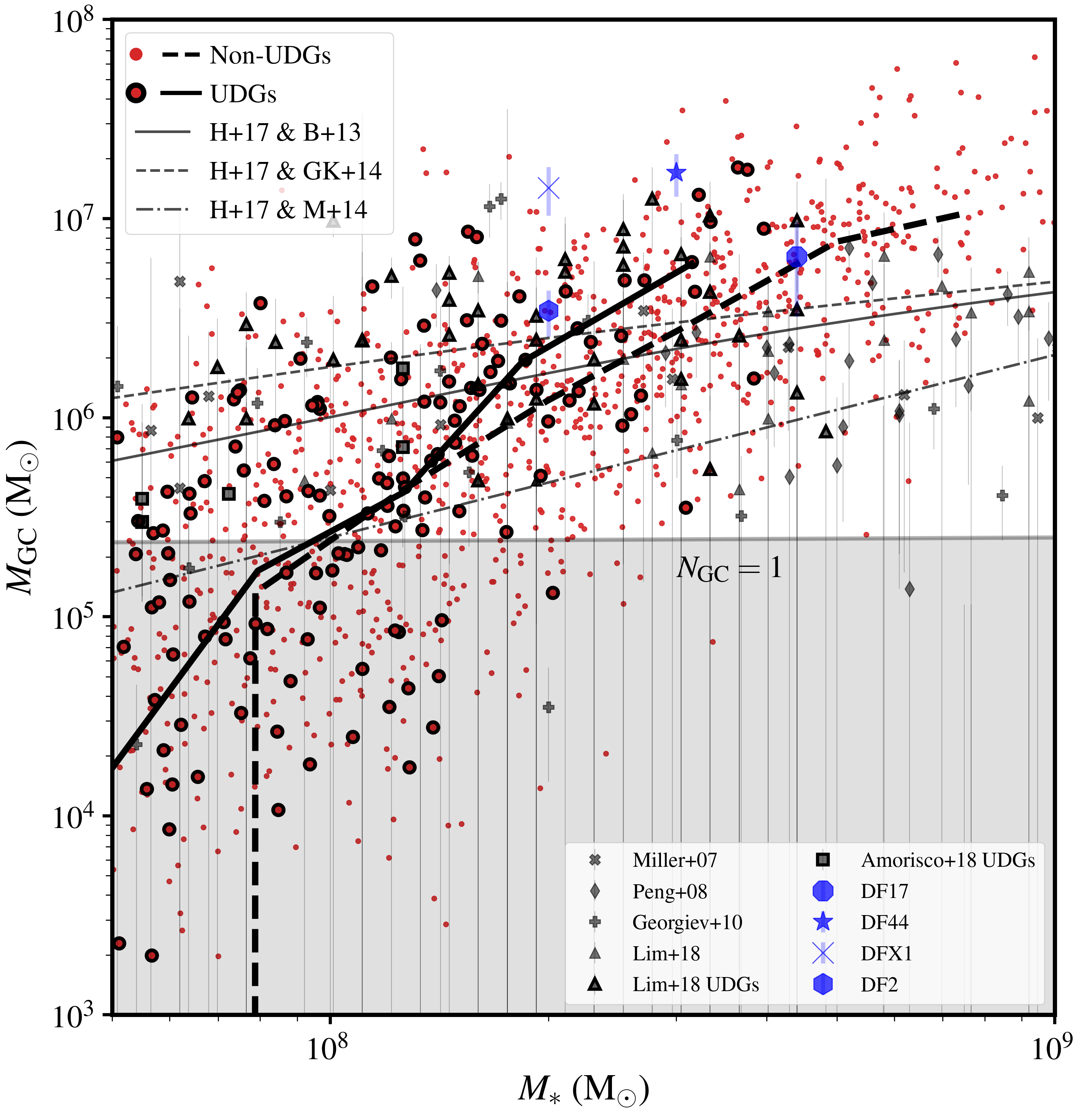}
	\caption{The relationship between stellar mass and globular cluster mass. Circular points are from our model, with 
		systems classified as UDGs highlighted with black outlines. The black dashed and solid lines show the median-binned trend for non-UDGs and UDGs respectively. Grey squares and triangles with black outlines show observations of GCs in UDGs from \protect \cite{amorisco2018} and \protect \cite{lim2018}, assuming an average mass of $2.3\scinot{5}~\msun{}$ \protect \citep{harris2017}. Grey ``x"s, diamonds, crosses, and traingles are non-UDG observations from \protect \cite{miller2007}, \protect \cite{peng2008}, \protect \cite{georgiev2010}, and \protect \cite{lim2018}. The blue octagon, pentagon, star, ``X", and hexagon highlight five notable UDGs of DF17, VCC 1287, DF44, DFX1, and DF2 respectively, from \protect \cite{peng2016}, \protect \cite{beasley2016}, \protect \cite{vandokkum2016}, \protect \cite{vandokkum2017} and \protect \cite{vandokkum2018}. The grey lines represent the expectations from a constant $M_{\rm GC}$-to-$M_{\rm halo}$ ratio of $2.9\scinot{-5}$ \protect \citep{harris2017}, following the abundance-matching relations of \protect \citealt{behroozi2013}  (solid grey line) \protect \citealt{garrison-kimmel2014} (dashed grey line) and \protect \citealt{miller2014} (dot-dashed line). The grey shaded region highlights where $M_{\rm GC}$ is less than the average mass of one GC from \protect \citep{harris2015}. }
	\label{fig:mgcvsmstar}
\end{figure}

While fully accounting for the disruption of globular clusters requires a more precise model for the birth of globular clusters and a higher resolution simulation, we model the effects of disruption on the globular cluster population to generate a more realistic GC population. Constraints from the distribution of GC ages and masses suggests two primary phases of globular cluster disruption \citep{fall2012}. First, while the cluster still resides in the disk, interactions with nearby molecular clouds are able to disrupt clusters at a roughly constant rate. Second, once the cluster is outside the disk, tidal interactions with the galaxy disk as well as multi-body interactions within the cluster cause the cluster to evaporate.

\begin{figure*}
	\centering
	\includegraphics[width=1\linewidth]{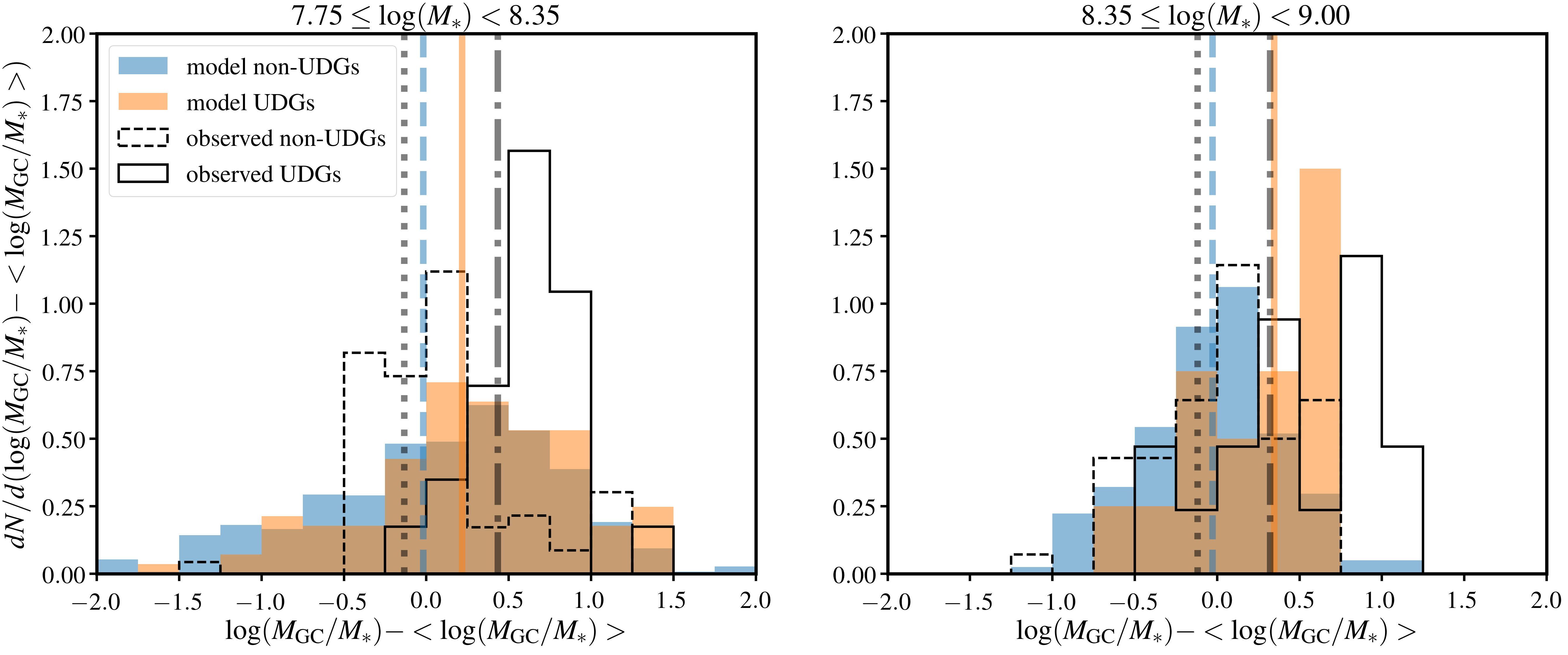}
	\caption{A comparison of the globular cluster populations of UDGs and non-UDGs among less massive ($10^{7.75}\le M_*/{\rm M_\odot}<10^{8.35}$; left) and more massive ($10^{8.35}\le M_*/{\rm M_\odot}<10^{9}$; right) objects. The blue and orange solid histograms show the GC mass fraction distributions from our model for a mass-matched sample of non-UDGs and UDGs respectively. To facilitate a comparison with observations, we have offset the distributions by the median GC mass fraction of all simulated objects in the appropriate mass range. Solid and dashed black histograms illustrate the mass-matched distributions for observed UDGs and non-UDGs from \protect \cite{miller2007}, \protect \cite{peng2008}, \protect \cite{georgiev2010}, \protect \cite{beasley2016}, \protect \cite{vandokkum2016}, \protect \cite{vandokkum2017}, \protect \cite{lim2018}, and \protect \cite{vandokkum2018} offset by the median \emph{observed} GC mass fraction for this sample. The solid orange and black lines show the median values for simulated and observed UDGs respectively, whereas the dashed blue and dotted black lines show the median values for non-UDGs. Our model predicts that massive UDGs have a $0.32$~dex higher GC mass fraction than non-UDGs, similar to observations finding an offset of $0.29$~dex. Less massive UDGs have a $0.18$~dex higher GC mass fraction than non-UDGs, compared with observations finding a $0.49$~dex offset.}
	\label{fig:deltahist}
\end{figure*}

To account for tidal {disruption} of GCs and cluster evaporation, we adopt equations $3$ and $4$ from \cite{gnedin2014}, using the combined stellar mass and dark-matter profiles to determine the orbital frequency. Each globular cluster is assumed to reside at its birth position for the duration of the simulation.\footnote{Within the central cores of the dark-matter halos (where nearly all GCs live), the GC disruption rate is not dependent on galacto-centric radius, so including the effects of dynamical friction would not significantly affect our results.}
The rate of disruption due to interactions with gas and stars in the disk (molecular clouds in particular) is less well constrained, particularly for galaxies with low stellar masses, although it has been shown to be related to the density of gas in the disk \citep{kruijssen2015}.
Rather than fully model the destruction of globular clusters in this phase, we parameterize the disruption rate as a linear function of the stellar and gas mass of the galaxy. In agreement with previous models (e.g.~\citealt{kruijssen2015}), we model the GC disruption rate as:
\begin{equation}
\frac{dM(t)}{dt}=M(t)/t_{\rm d,~disk},
\end{equation}
where $M(t)$ is the globular cluster mass at simulation time $t$, and $t_{\rm d,~disk}$ is the disruption timescale due to the fact that the cluster is born in the disk. We parameterize $t_{\rm d,~disk}$ as a function of the stellar mass and gas fraction of the galaxy in the simulation as:
\begin{equation}
\log{t_{\rm d,~disk}}=C_1\log\left(\frac{M_{\rm gas}}{M_*}\right)+C_2\log\left(\frac{ M_{*}}{\rm M_\odot}\right)+C_3,
\end{equation}
where $M_{\rm gas}$ is the gas mass of the object, and $M_*$ is the stellar mass of the object. The constants $C_1$, $C_2$, and $C_3$ are determined to be the best fit values to reproduce the observed ratio of GC-to-stellar mass for three bins of $M_*$ from $10^8$~\msun{} to $10^9$~\msun{}. The best fit values are $C_1=-0.4$, $C_2=0$, and $C_3=0.33$.

{Additionally, we model the effects of tidal stripping of GCs by the cluster environment. To incorporate this effect into our model, we make the assumption that the amount of GC mass stripped is equivalent to the amount of stellar mass stripped: 
	\begin{equation}
	M_{\rm GC,~stripped}/M_{\rm GC,~infall}=M_{*,~{\rm stripped}}/M_{*,~{\rm infall}},
	\label{eqn:stripeqn}
	\end{equation}
	where $M_{*,~{\rm stripped}}/M_{*,~{\rm infall}}$ is taken from the same \cite{errani2018} tracks used to model the stellar mass stripping of the galaxy. This is motivated by observations indicating that the extent of globular cluster populations around UDGs are similar to the extent of the stellar disk \citep{amorisco2018}.
To test the impact of this assumption on our analysis, we run an alternative model using star particles in the simulation as tracer particles \citep[in a method similar to][]{mistani2016}. For each GC that is born, we tag a random star particle as representing that GC at its birth. If the star particle is not bound to the galaxy at $z=0$, the GC is considered to be stripped. This procedure only changes the fraction of stripped GCs by $<10\%$ (with our fiduciary model stripping $10\%$ more GCs than the alternative model), so we conclude that the assumptions of our stripping model don’t significantly change our conclusions. {Additionally, we explore how the initial distribution of GCs affects the stripping of GCs and the GC distribution in Appendix~\ref{sec:appendix}. We find that in the Illustiris-TNG simulation, particles that start off with a compact distribution (like our modeled GCs) follow Equation~\ref{eqn:stripeqn} very well, regardless of the exact details of the distribution. We also show that GCs with more compact distributions expand significantly due to tidal heating, so that our assumption that the GC distribution is similar to the stellar distribution at infall is consistent with observations that the GC distribution is slightly more extended than the stellar distribution in GCs \citep{amorisco2018}.} Furthermore, tidal stripping by the cluster is a significantly weaker effect than GC disruption in our model, so if the GC population is taken to be significantly more extended, the GC mass actually increases because of the less effective GC disruption. Lastly, we note that while increasing the extent of the GC population weakens the trend between total GC mass and infall time (resulting in a smaller difference between UDGs and non-UDGs), it also weakens the trend between GC abundance and environment (in contrast with existing observations implying a significant environmental dependence to GC abundances, e.g.~\citealt{peng2008}).}

Following this procedure, our analysis generates a population of galaxies affected by tidal heating and stripping, as well as their globular-cluster populations. Figure~\ref{fig:fracgcvszpeakbymass812m1} shows the relationship between the total GC mass-to-stellar mass ($M_{\rm GC}$-to-$M_*$) ratio and infall redshift (throughout this analysis, we will focus primarily on the total GC mass formed in our model systems because that is less sensitive to the uncertain nature of GC disruption than the abundance of GCs). This figure also shows how models without disruption produce GC-to-stellar mass ratios of $10-50\%$, far beyond what is observed.

In agreement with \cite{mistani2016} (as well as simulations from \citealt{pfeffer2018}), we find a strong correlation between GC fraction and infall time. Although this effect is dampened by stripping and disruption of globular clusters, this correlation persists, particularly for systems with infall redshift above $2$. {On average, we find $\log\left(M_{\rm GC}/M_*\right) \propto (0.1\pm0.02)z_{\rm inf}$ for massive systems and $\log\left(M_{\rm GC}/M_*\right) \propto (0.22\pm0.03)z_{\rm inf}$ for less massive systems}. Figure~\ref{fig:mgcvsmstar} compares our models and observations in the $M_{\rm GC}-M_*$ plane. Our model systems approximately line up with observations for objects $M_*\ge 10^{7.75}$~\msun{}. While limitations in our modeling (see Sec.~\ref{sec:modellimitations}) prevent an exact match between observations and model points, our models capture the general trend of globular clusters that are continually formed in regions of intense star formation and disrupted. {Furthermore, it is clear that UDGs (which have earlier infall times than non-UDGs -- see Fig.~\ref{fig:fracgcvszpeakbymass812m1}) have significantly higher globular cluster masses than non-UDGs of similar stellar mass.}

\section{Comparing GC Populations of UDGs and non-UDGs}
\label{sec:results}
{A key prediction of the \cite{carleton2019} model is that systems with the earliest infall times are most likely to evolve into {cluster} UDGs (given the greater accumulated tidal effects). As such, objects identified as UDGs (defined as systems with stellar surface density $\Sigma_*=M_*/(\pi r_e^2)$ between $1.73\times 10^6$ and
$17.3\times 10^6$~\msun{}~kpc$^{-2}$ and $r_e$ between $1.5$ and $7$~kpc) tend to have higher $M_{\rm GC}$ values than non-UDG dwarfs at a similar mass. To highlight this difference in particular, we compare the GC mass fraction ($M_{\rm GC}/M_*$) distributions of UDGs and a stellar mass-matched sample of non-UDG dwarf galaxies in Figure~\ref{fig:deltahist}. We find that UDGs have a $0.32$~dex higher GC fraction than non-UDGs among the most massive systems (right panel).}
\begin{figure}
	\centering
	\includegraphics[width=1\linewidth]{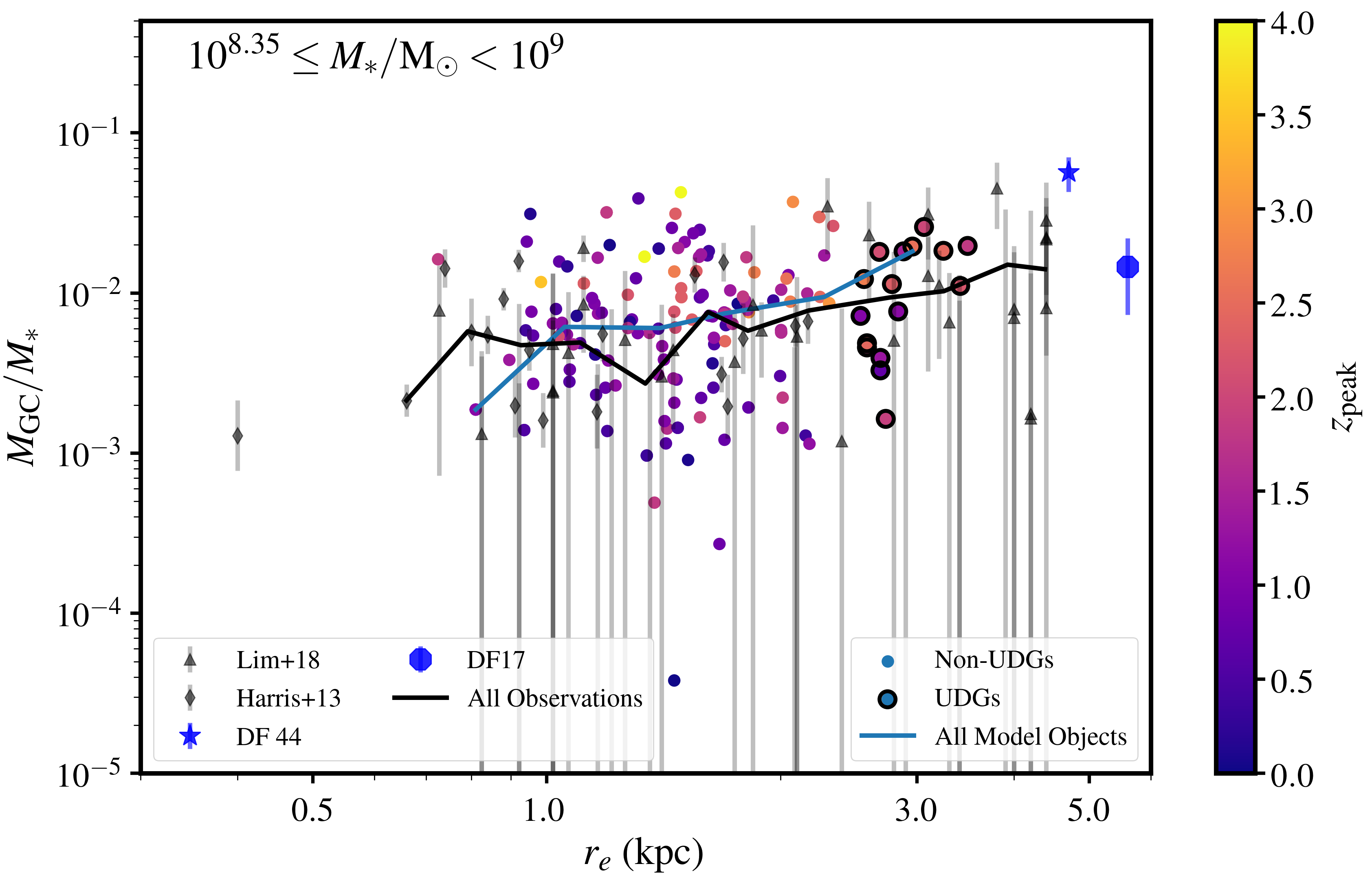}
	\caption{The relationship between GC mass fraction and size among a mass-matched sample of UDG and non-UDG model systems with $M_*\ge10^{8.35}$~\msun{}, as well as observed systems. Circular points are from our model, color-coded by $z_{\rm peak}$ (the redshift of the objects maximum halo mass), and systems classified as UDGs are highlighted with black outlines. Grey triangles are observations from \protect \cite{lim2018} and grey diamonds are from the \protect \cite{harris2013} collection of observations. As in Fig.~\ref{fig:mgcvsmstar}, the blue star refers to DF44 \protect \citep{vandokkum2016} and the blue octogon refers to DF17 \protect \citep{peng2016}. The blue line shows the median GC fraction among all model objects (UDGs and non-UDGs) in bins of $r_e$, and the black line shows the binned median points considering all observations. Model objects show a positive trend between size and GC fraction, both of which are correlated with infall redshift. This is consistent with the strong trend observed between size and globular cluster fraction in massive dwarf galaxies.}
	\label{fig:gcvssize}
\end{figure}
\begin{figure}
	\centering
	\includegraphics[width=1\linewidth]{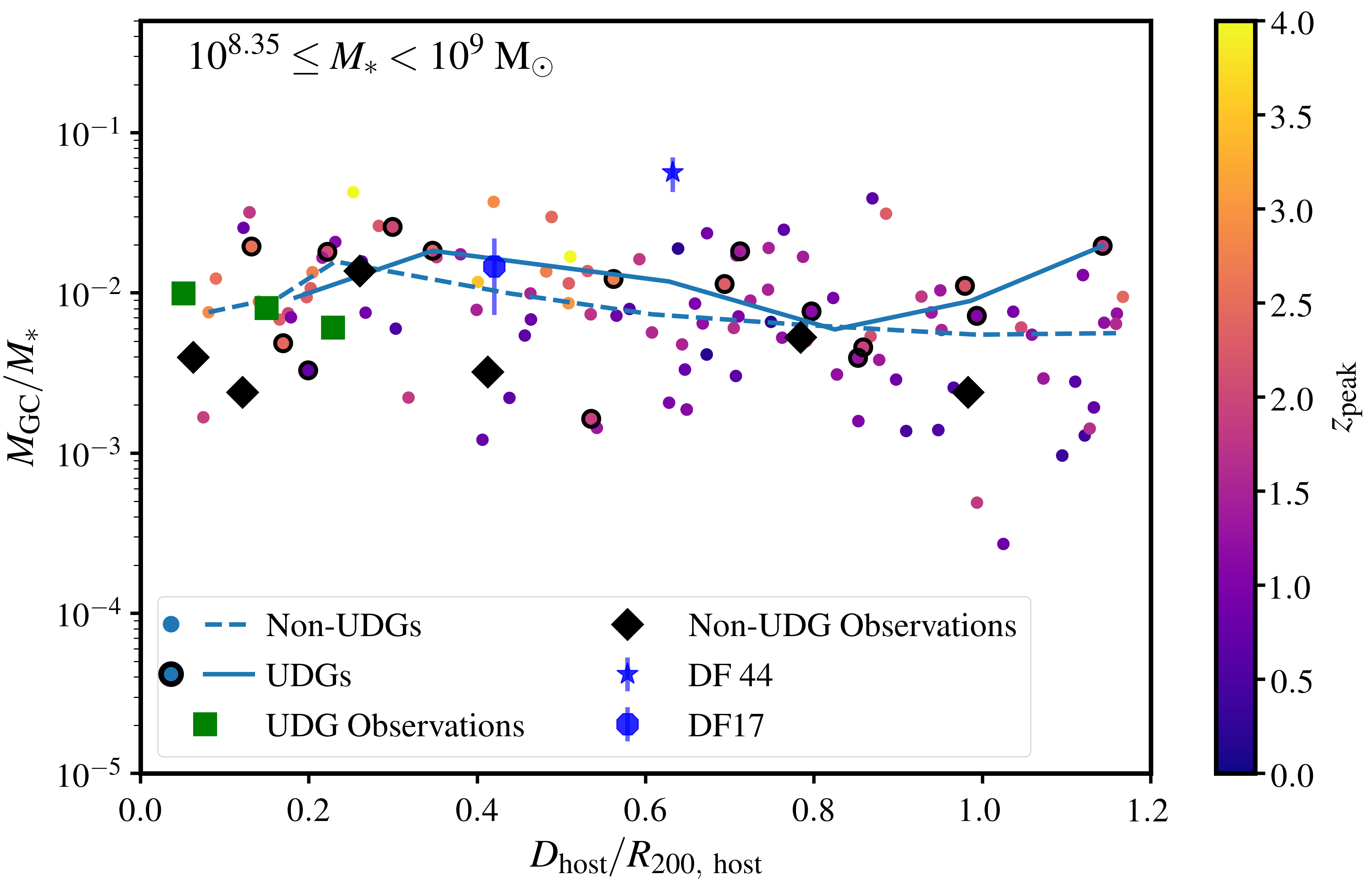}
	\caption{The relationship between globular cluster fraction and cluster-centric distance (normalized to $R_{200}$ of the host cluster). The relative frequency of globular clusters increases as systems approach the cluster center because early infalling dwarfs with larger globular cluster populations are able to sink toward the cluster center. The model symbols are the same as in Fig.~\ref{fig:gcvssize}. Observations of a large number of systems in the cluster outskirts are difficult to obtain, but we show the binned median GC fraction for observed UDGs (green squares) and non-UDGs (black diamonds). Future observations of GCs around UDGs in the cluster outskirts can further test the predicted environmental dependence of GC mass among UDGs.}
	\label{fig:gcfracdhost}
\end{figure}
\begin{figure*}
	\centering
	\begin{tabular}{cc}
		\includegraphics[width=.5\linewidth]{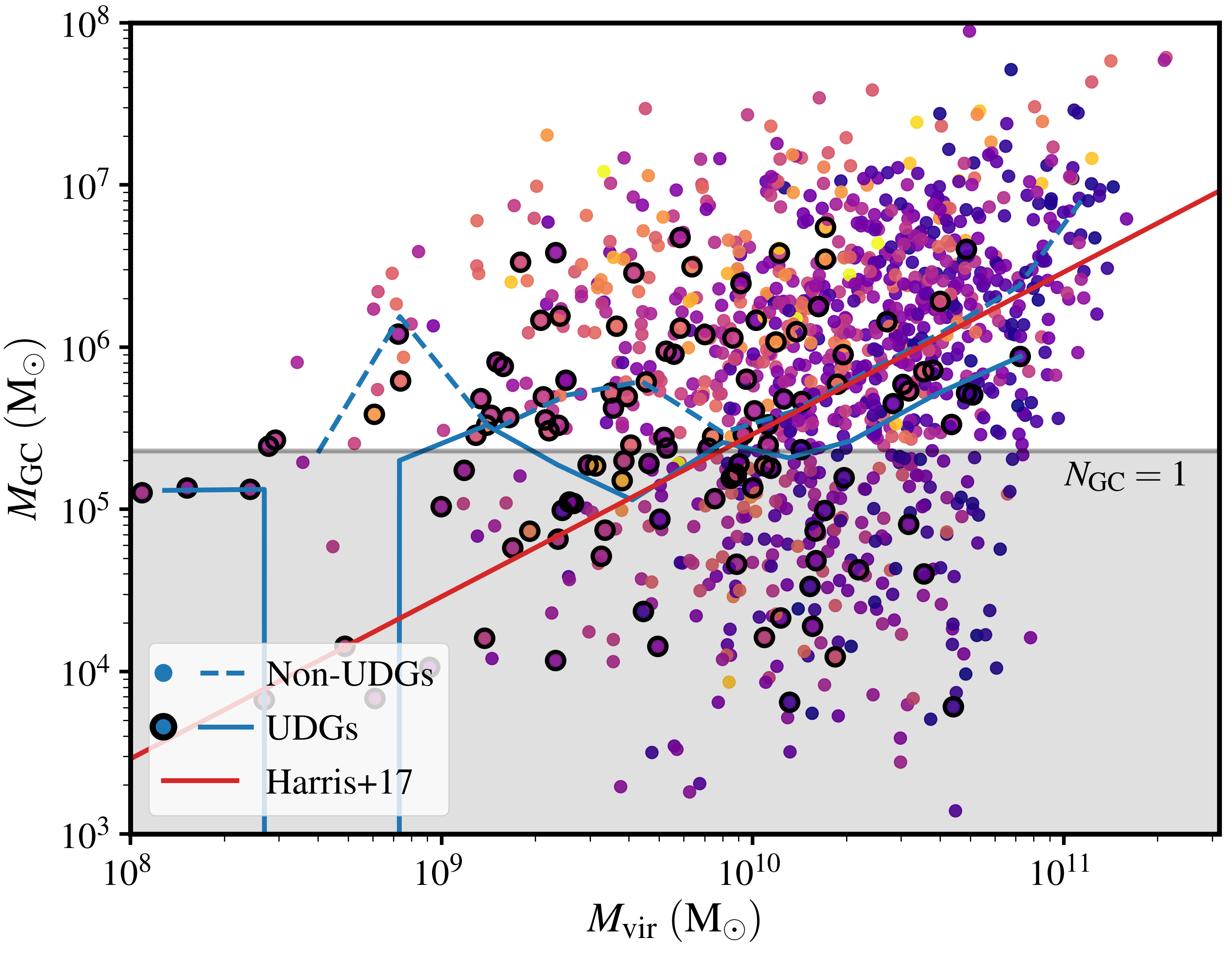}
		\includegraphics[width=.5\linewidth]{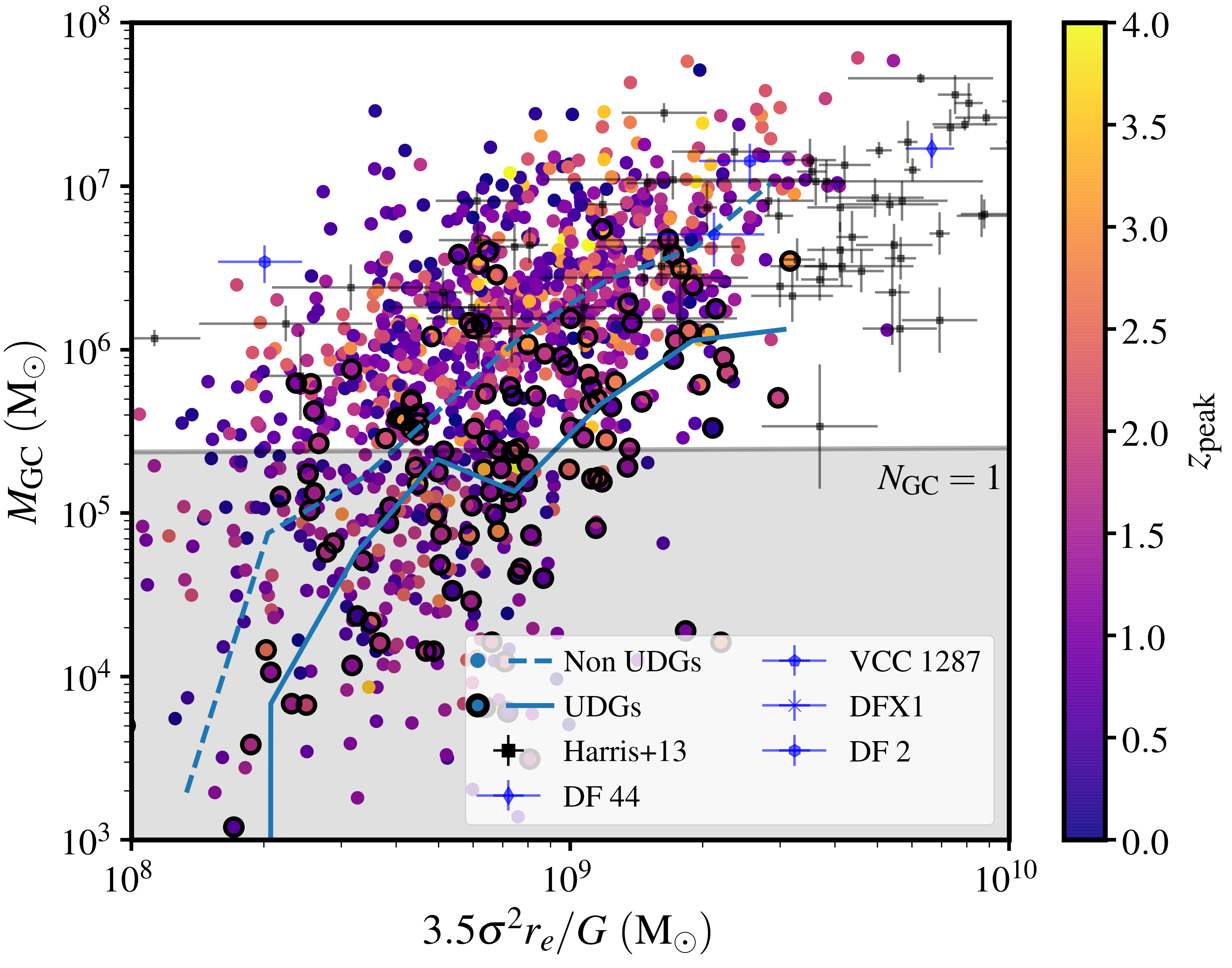} 
	\end{tabular}
	\caption{{\bf Left:} {The relationship between GC mass and halo mass for objects with $M_*>10^{7.75}~\msun{}$}. Symbols are the same as in Fig.~\ref{fig:gcvssize}, and the red line is the constant GC-to-halo mass ratio from \protect \cite{harris2017}. Non-UDGs in our model reproduce the GC-to-halo mass ratio for massive objects. However, UDGs have a slightly lower GC-to-halo mass ratio. {\bf Right:} The relationship between total globular cluster mass and dynamical mass (as probed by stellar velocity dispersion) for dwarf galaxies. Because UDGs have a higher dynamical mass than non-UDGs of similar stellar mass, there is an offset between UDGs and non-UDGs in this space among our models that may be seen in future observations.}
	\label{fig:mgcvssigma}
\end{figure*}
The main cause of this difference in GC populations is the offset infall time distributions between UDGs and non-UDGs. {Any model in which UDGs have significantly earlier infall times than non-UDGs \citep[e.g.~][]{yozin2015,tremmel2019} should produce an elevated number of GCs in UDGs compared with non-UDGs.}
A smaller difference between UDGs and non-UDGs is predicted in less massive systems (left panel).
{This difference is not caused by the slight difference between GC formation and disruption among high- and low-mass dwarfs (in fact, because disruption is more efficient in high-mass dwarfs, there is a stronger correlation between GC fraction and infall time for low-mass dwarfs than high-mass dwarfs)}; rather, a larger fraction of low-mass UDGs have later infall times (Fig.~\ref{fig:fracgcvszpeakbymass812m1}). This is because lower mass dwarfs only require $\sim50\%$ expansion to be considered a UDG, whereas higher mass dwarfs must expand by a factor of $\sim2$ based on our assumed mass-size relation. If galaxies are larger at infall (as suggested by simulations: e.g.~\citealt{genel2018,tremmel2019,jiang2019,wright2020}) less expansion is required, so the difference in GC populations may be less significant.

To further compare our model with observations, Figure~\ref{fig:gcvssize} shows the relationship between globular cluster mass fraction and galaxy half-light radius for objects with $M_*\ge 10^{8.35}$~\msun{}.
As both tidal expansion and globular cluster mass fraction are correlated with infall time, our model predicts a positive correlation between GC fraction and size among UDGs and non-UDGs (Spearman rank correlation $p$ value is $0.008$). Among model points with $M_{\rm GC}>0$, we find $\log\left(M_{GC}/M_*\right)\propto (0.5\pm0.2)\log\left(r_e\right)$, which is very similar to observations showing $\log\left(M_{GC}/M_*\right)\propto (0.6\pm0.2)\log\left(r_e\right)$. That this correlation persists for all $r_e$ values highlights the fact that UDGs are not unique in this model, but simply the high-size tail of the galaxy population.
Similarly, Figure~\ref{fig:gcfracdhost} shows the correlation between GC fraction and cluster-centric distance. A slight negative correlation between globular cluster fraction and cluster-centric distance is present, again resulting from the correlations between infall time, cluster-centric distance, and GC fraction. This trend is consistent between UDGs and non-UDGs and is similar to the trend observed among non-UDGs in Virgo \citep{peng2008}. Observations of UDGs have not yet been able to probe a large range of cluster-centric distances, but our predicted trend is consistent with a slight negative correlation between GC mass fraction and cluster-centric distance among existing observations of UDGs \citep{lim2018}. We predict that a similar trend will be present among UDGs and non-UDGs once objects in the cluster outskirts can be studied in more detail. Specifically, we find  $\log\left(M_{\rm GC}/M_*\right)\propto (-0.30\pm0.06)\log\left(D_{\rm host}/R_{200}\right)$.

{Given the constant ratio of GC mass-to-dark-matter mass ratio observed among massive objects \citep{harris2017}, GC populations have been particularly intriguing as a window to the dark-matter mass of galaxies. In Figure~\ref{fig:mgcvssigma}, we plot globular cluster mass vs. both total halo mass (left) and dynamical mass as probed by stellar velocity dispersion (right).}
{Stellar velocity dispersions are determined using the line-of-sight Virial Theorem assuming a Plummer stellar distribution and the stripped dark-matter density profile, as in \cite{carleton2019}.}
Notably, our model is within $0.1$~dex of the $M_{\rm GC}/M_{\rm halo}$ ratio observed in massive galaxies \citep{harris2017} for objects with $M_{\rm halo}>10^{10}$~\msun{}, despite our use of observations of the $M_{\rm GC}/M_*$ ratio, not the $M_{\rm GC}/M_{\rm halo}$ ratio, to constrain our model. However, a large degree ($\sim1$~dex) of scatter is present, compared with the $0.26$~dex scatter seen observationally \citep{harris2017}. {Additionally, UDGs have systematically \emph{lower} globular cluster masses at a given halo mass (by $0.4$~dex; left panel) and dynamical mass (by $0.8$~dex; right panel).}
There are two reasons for the offsets between UDGs and non-UDGs in this space. Firstly, UDGs with the largest halo masses are selected to have the latest infall times (because they are relatively unaffected by stripping), and thus lower GC masses (Fig.~\ref{fig:fracgcvszpeakbymass812m1}). Secondly, the tidal heating process increases an object's dynamical mass: as the stellar extent of a system increases substantially, dark-matter is preferentially stripped from the halo outskirts, leaving the central region probed by the stellar velocity dispersion largely intact. This results in higher dynamical masses for UDGs compared with non-UDGs \citep{carleton2019,errani2018}.
This prediction is in marginal conflict with a small number of observations indicating that UDGs and non-UDGs have a similar GC mass for a given dynamical mass. Future observations with large samples of UDGs are needed to explore the full range of dynamical masses for a given GC mass.
While this suggests that GC abundances may be better at estimating total halo mass than stellar dynamics, our modeling relies on more assumptions (a constant cluster formation fraction, a constant GC initial mass function, a constant GC radial distribution, and an unevolving dwarf-elliptical size-mass relation) than dynamical probes, and our models have difficulty in reproducing the GC populations of lower mass dwarfs. Nevertheless, this result \emph{does} show that a constant GC-to-halo mass ratio does not necessarily preclude a continuous GC formation process. An important caveat to this result is that a significant correlation between GC-to-halo mass ratio and infall redshift is present that could systematically affect halo mass inferences from GC abundances. Lastly, we note that the presence of systems like DF44, with a high GC mass for its dynamical mass, suggests that some UDGs may have a different formation path.

\begin{figure}
	\centering
	\includegraphics[width=1\linewidth]{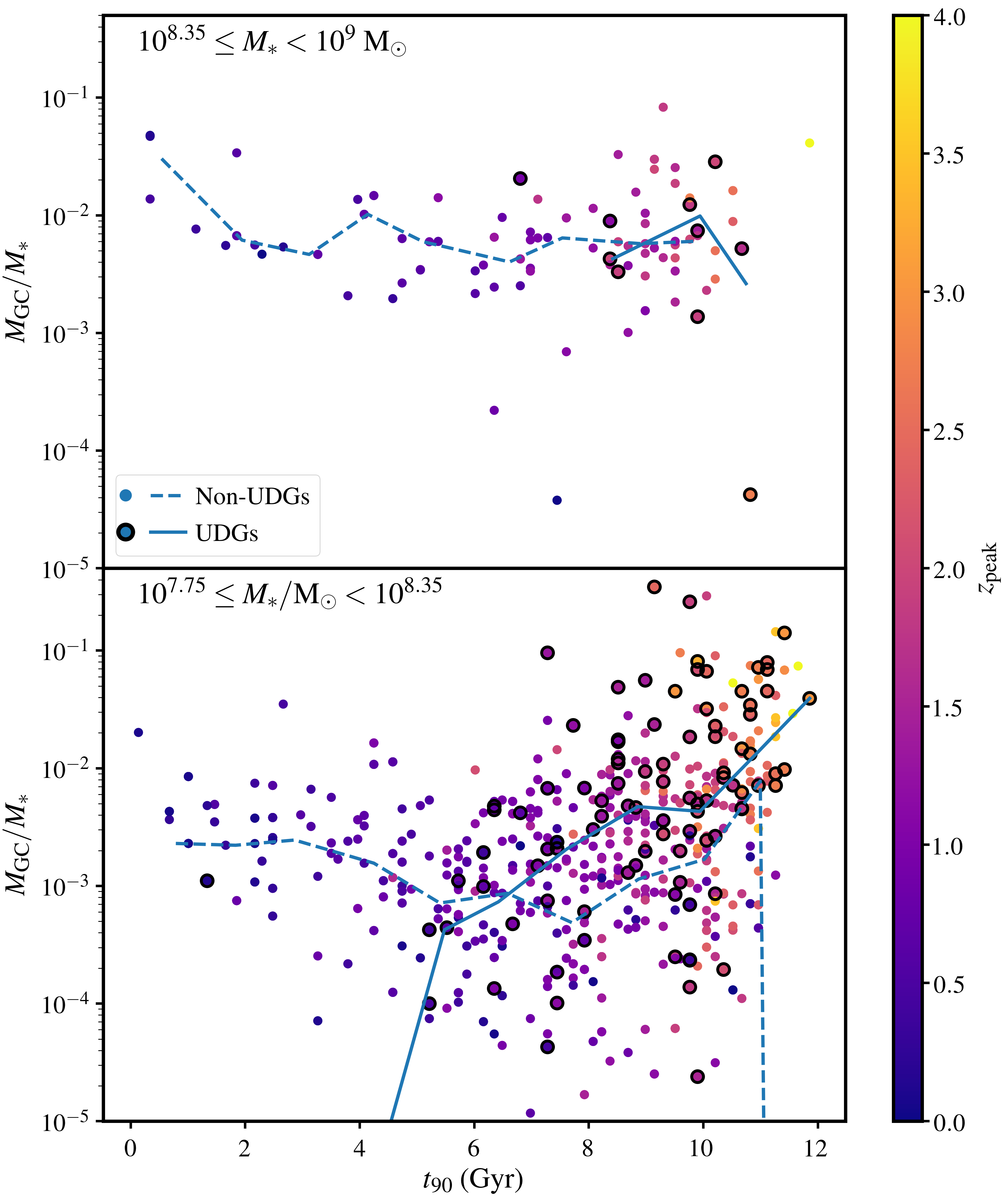}
	\caption{The correlation between GC fraction and stellar age (as defined by the lookback time for which $90\%$ of stars are formed). Both UDGs and non-UDGs with early ($t_{90}>7$~Gyr) quenching times have significantly higher GC fractions than later infalling systems.}
	\label{fig:fracgcvsagebymass812}
\end{figure}

A specific consequence of this model is a correlation between satellite infall time and globular cluster abundance (Fig.~\ref{fig:fracgcvszpeakbymass812m1}). Observationally, this manifests as a trend between GC fraction and stellar age (characterized by $t_{90}$: the lookback time at which a galaxy's stellar mass first reaches $90\%$ of its peak stellar mass), which is shown in Fig.~\ref{fig:fracgcvsagebymass812} for objects in two stellar mass bins. Although GC stripping and destruction substantially dampen the trend between stellar age and GC mass fraction for massive objects and objects with younger ages, a correlation between GC mass fraction and stellar age is present among less massive objects that should be detectable in future observations.
Among objects with $t_{90}$ greater than $7$~Gyr and $M_{\rm GC}>0$, we find $\log\left(M_{\rm GC}/M_*\right)\propto (1.1\pm0.7)\log\left(t_{90}/{\rm Gyr}\right)$ for objects with $M_*\ge 10^{8.35}$~\msun{} (Fig.~\ref{fig:fracgcvsagebymass812} upper panel) and $\log\left(M_{\rm GC}/M_*\right)\propto (3.3\pm0.3)\log\left(t_{90}/{\rm Gyr}\right)$ for objects with $10^{7.75}\le M_*< 10^{8.35}$~\msun{} (Fig.~\ref{fig:fracgcvsagebymass812} lower panel). Again, this trend is roughly the same between UDGs and non-UDGs; however, there is a $\sim0.5$~dex offset among the oldest systems. This offset is because of systems affected by early pre-processing (note objects with late $z_{\rm peak}$ values but early $t_{90}$ values in Fig.~\ref{fig:fracgcvsagebymass812}). In our model, systems that fall into a group at high $z$ have lower GC formation rates after falling into the group because of the decreased SFR densities. However, some gas is still present in the disk, so GCs are still disrupted at a rapid rate, resulting in lower GC masses at $z=0$.

\begin{figure}
	\centering
	\includegraphics[width=1\linewidth]{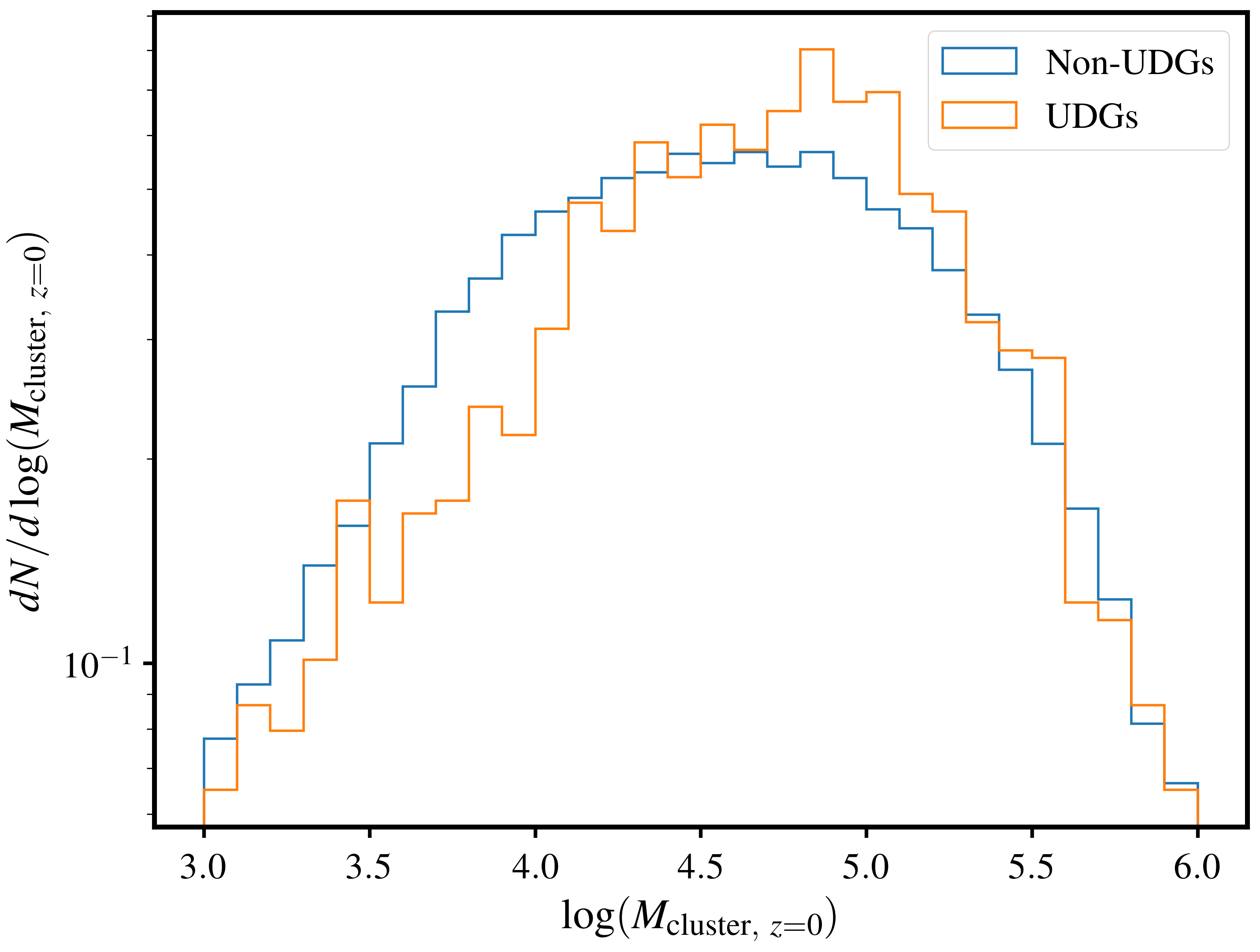}
	\caption{The stacked $z=0$ GC mass function for both non-UDGs (blue) and UDGs (orange). Because UDGs form their GCs at a higher redshift than non-UDGs, more low-mass GCs are destroyed by disruptive disk and tidal effects, resulting in a more top-heavy GC mass function.}
	\label{fig:stackhist}
\end{figure}

Another observable consequence of this model is that objects with early infall times have more top-heavy GC mass functions at $z=0$, as illustrated in the stacked GC mass functions in Fig.~\ref{fig:stackhist}. This is because the less massive GCs are not able to survive many Gyr without evaporating. Specifically, systems with infall redshifts $>2$ have a $0.44$~dex higher mean GC mass than systems with later infall times. As UDGs have preferentially early infall times, they have a $0.13$~dex higher mean GC mass than non-UDGs. This offset is important to note, particularly as observations typically assume a constant GC luminosity function to correct for incompleteness among observations of lower mass GCs \citep[e.g.~][]{lim2018}.
This also illustrates how a UDG like DF2 could have a very top heavy mass function, and should be testable in future observations. Similarly, GCs in early infall systems have older stellar ages than those in later infall systems, but this difference ($\sim11$~Gyr old vs. $\sim8$~Gyr old) may be difficult to distinguish observationally.

\section{GCs in Low-Mass Galaxies}
\label{sec:modellimitations}
Our model predicts that GCs should be very sparse in the lowest mass ($M_*<10^{7.75}$~\msun{}) galaxies. This is in contrast with observations indicating that GCs are present in most systems above $10^7$~\msun{}, pointing to an incompleteness in our model.
This discrepancy is magnified because low mass galaxies only have  a handful of GCs --- a few additional GCs can double the globular cluster abundance among these systems.
Illustris-TNG is limited by temporal and spatial resolution at very high redshifts, so primordial GC formation is largely excluded from our analysis. The presence of a small number of primordial GCs in low mass galaxies would be enough to resolve the discrepancy between our model points and observations while preserving our results for higher mass objects. Additionally, assumptions in our modeling of GC disruption in low-mass objects could affect our ability to accurately characterize the GC population around these systems. For example, it is possible that globular clusters in low mass galaxies are able to escape to larger galacto-centric radii (given the weaker gravitational pull of the galaxy), and become less susceptible to disruption.
Altogether, while our model is effective for high-mass dwarfs, illustrating the effect of infall time on total GC mass, these limitations prevent an accurate modeling of GCs around low mass dwarfs.

\section{Conclusion}
\label{sec:conclusions}
We combine models for globular cluster and UDG formation to investigate the expected GC populations of tidally-heated UDGs {in clusters}. {Two key features of our model are that GCs are formed more efficiently at high redshift and UDGs have preferentially early infall times. Combined, these features suggest that UDGs in our model have higher GC abundances than non-UDGs of similar mass.}
\begin{itemize}
	\item Observations finding that UDGs have a higher GC mass than non-UDGs can be explained in a scenario in which GCs are formed during periods of intense star formation at high $z$ and UDGs are formed through tidal heating of normal dwarfs with early cluster infall times. 
	\item This model predicts that a galaxy's GC mass fraction should be correlated with its stellar age, half-light radius, and cluster-centric distance.
	\item This model also predicts that UDGs should have a more top-heavy GC mass function than non-UDGs (in qualitative agreement with one observation), so extrapolating GC abundances based on a non-UDG GC mass function may overestimate the true GC abundance.
	\item A correlation between the GC-to-halo mass ratio of dwarf galaxies and their infall time is expected, and UDGs are predicted to have lower GC-to-halo mass ratios and lower GC-to-dynamical mass ratios than non-UDGs.
\end{itemize}

\section*{Acknowledgments}
The authors are grateful to the anonymous referee for their helpful comments. TMC is grateful for Manoj Kaplinghat for his helpful comments. Additionally, the authors are grateful to the anonymous reviewer for their constructive comments.
This research made use of {\texttt{Astropy}}, a community-developed core Python
package for Astronomy \citep{astropy}. Additionally, the Python packages
{\texttt{NumPy}} \citep{numpy}, {\texttt{iPython}} \citep{ipython},
{\texttt{SciPy}} \citep{scipy}, and {\texttt{matplotlib}} \citep{matplotlib}
were utilized for the majority of our data analysis and presentation.

\section{Data Availability}
The data underlying this article will be shared on reasonable request to the corresponding author.

\bibliography{udggcpaper}

\appendix
\section{Effect of GC Radial Distribution on GC Stripping}
{
\label{sec:appendix}
Here, we explore how different assumed GC radial distributions affect the amount of GC stripping. In our fiducial scenario, stripping of GCs is less important than GC disruption. However, our results are sensitive to the initial distribution of GCs around dwarfs. In massive galaxies, many GCs are accreted into the halo, resulting in a very extended GC distribution \citep[e.g.][]{mackey2004,beasley2018}. For UDGs, however, the GC distribution is very similar to (although slightly more extended than) the distribution of the stellar disk \citep{amorisco2018}. Indeed, if a significantly more extended GC population is assumed\footnote{For this case, GC particles are distributed following a $r^{-3.5}$ distribution from $0$ to $10$~kpc at birth, and their positions are tracked as the simulation evolves.}, the trend between GC fraction and galaxy size (Fig.~\ref{fig:gcvssize}) goes away. However, in this case, the $z=0$ GC distribution is more extended than observations indicate. Notably, the GC distribution becomes \emph{more extended} at $z=0$ than at their birth --- tidal heating puffs up the GC distribution. 

To further test how the initial GC distribution affects our results, we track tracer particles following a particular spatial distribution around galaxies in Illustris-TNG. In particular, at $z_{\rm peak}$, $100$ particles (either star or dark-matter particles) are drawn from the simulation, such that they follow a power law distribution, with power-law exponent $p$ and half-light radius $r_p$ as a fixed fraction of the initial stellar half-light radius. The model parameters are illustrated in Table~\ref{tab:modeltable}. We then measure what fraction of the initial particles are within the halo virial radius at $z=0$ and what the final half-number radius is.

\begin{figure}
	\centering
	\includegraphics[width=1\linewidth]{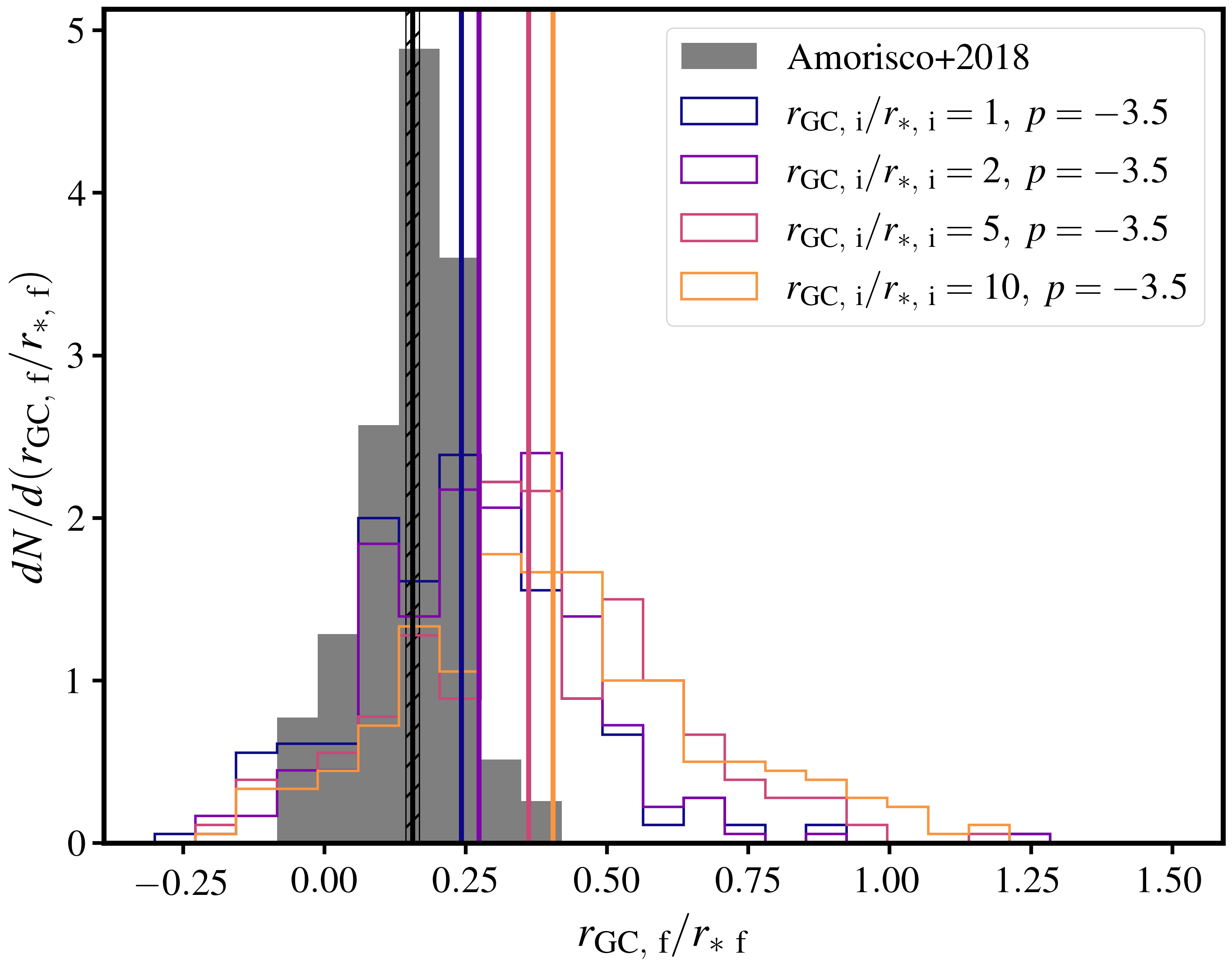}
	\caption{The distribution of $r_{\rm GC}/r_*$ at $z=0$ for GC distributions following a power law of $p=-3.5$ and various $r_{GC}/r_*$ values at infall. Also shown are the observed $r_{\rm GC}/r_*$ distributions from \protect \cite{amorisco2018} as the grey histogram. The vertical lines correspond to the mean of the distributions.}
	\label{fig:rrgcudg}
\end{figure}

Figure~\ref{fig:rrgcudg} illustrates the distribution of $r_{\rm GC}/r_*$ among our sample, compared with observations. Because most GC distributions become more \emph{extended} due to tidal heating, the GC distribution at infall must be more compact than the $z=0$ observations. Among our models, the most compact GC distribution best matches the observed GC distributions around UDGs in Coma from \cite{amorisco2018}. This validates our assumption that the GC distribution is similar to the stellar distribution.

\begin{figure}
	\centering
	\includegraphics[width=1\linewidth]{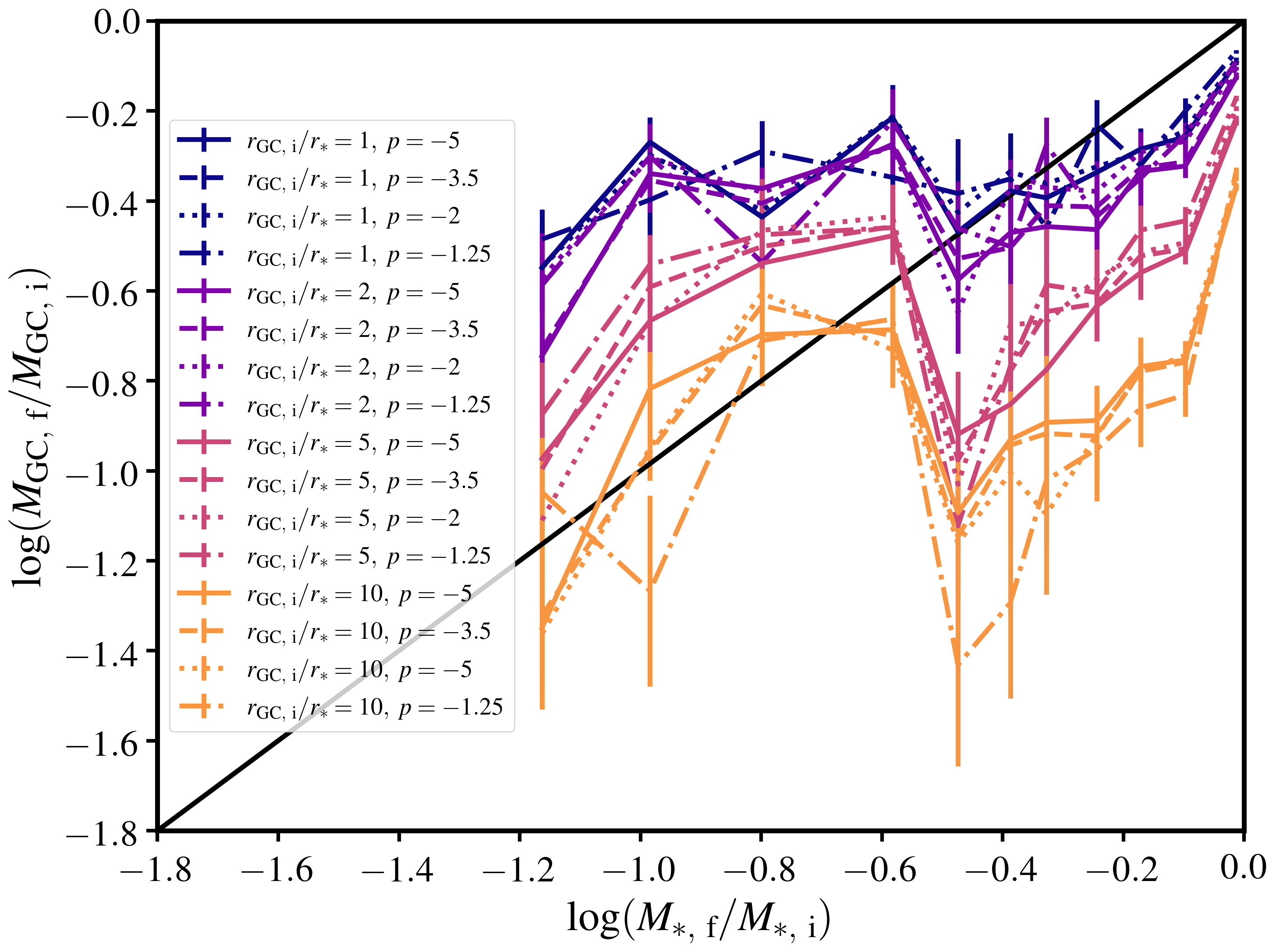}
	\caption{The relationship between GC mass loss and stellar mass loss in our analysis of GC stripping. Regardless of the initial GC distribution, stripped GC mass and stripped stellar mass are highly correlated, and if the initial GC distribution follows a power law that is more compact than a few times the stellar distribution, it loses a similar amount of mass compared with the stellar population. }
	\label{fig:mstlossmlossgc}
\end{figure}

Figure~\ref{fig:mstlossmlossgc} shows the correlation between the amount of stellar mass stripped and the amount of GC mass stripped in this experiment. Clearly, the amount of GC stripping is not dependent on the slope of the GC distribution. Also, for models with $r_{\rm GC}\simeq r_*$ and moderate stellar stripping, the amount of GC stripping is very similar to stellar distribution. For models with more extreme stellar stripping, the amount of GC stripping is \emph{less} than the amount of stellar stripping. This analysis, combined with the fact that $r_{\rm GC}/r_*=1$ at infall best matches observations, suggests that our assumption of $M_{\rm GC,~stripped}/M_{\rm GC,~infall}=M_{*,~{\rm stripped}}/M_{*,~{\rm infall}}$ is reasonable.

\begin{table}
		\label{tab:modeltable}	
	\caption{The parameters of our models studying the effect that the initial GC distribution has on GC stripping.}
	\begin{tabular}{|c|c|c|}
	\hline 
	Model & Power Law Exponent $p$ & $r_{{\rm GC},~i}/r_{*,~f}$ \\ 
	\hline 
	A & $-5$ & $1$ \\ 
	B & $-5$ & $2$ \\ 
	C & $-5$ & $5$ \\ 
	D & $-5$ & $10$ \\ 
	E & $-3.5$ & $1$ \\ 
	F & $-3.5$ & $2$ \\ 
	G & $-3.5$ & $5$ \\ 
	H & $-3.5$ & $10$ \\ 
	I & $-2$ & $1$ \\ 
	J & $-2$ & $2$ \\ 
	K & $-2$ & $5$ \\ 
	L & $-2$ & $10$ \\ 
	M & $-1$ & $1$ \\ 
	N & $-1$ & $2$ \\ 
	O & $-1$ & $5$ \\ 
	P & $-1$ & $10$ \\ 					
	\hline 

	\end{tabular} 
\end{table}
}

\end{document}